\documentclass[preprint,prc,showpacs,preprintnumbers,amsmath,amssymb]{revtex4-2}
\usepackage{graphicx,latexsym,amssymb,amsmath,color,multirow,mathrsfs,booktabs,hyperref,ifpdf}
\usepackage{dcolumn}% Align table columns on decimal point
\usepackage{bm}% bold math
\usepackage{longtable}
\def\pg{$(p,\gamma)$\ }

\begin{document}
\title{Bound-to-continuum potential model for the $(p,\gamma)$ reactions of 
the CNO cycle}
\author{Nguyen Le Anh$^{1,2,3}$}
\email{anhnl@hcmue.edu.vn}
\author{Bui Minh Loc$^{4,5}$}
\email{buiminhloc@tdtu.edu.vn (Corresponding author)}
\affiliation{{$^1$} Department of Theoretical Physics, Faculty of Physics and Engineering Physics, University of Science, Ho Chi Minh City, Vietnam.
\\
{$^2$} Vietnam National University, Ho Chi Minh City, Vietnam.
\\
{$^3$} Department of Physics, Ho Chi Minh City University of 
Education\\280 An Duong Vuong, Ward 4, District 5, Ho Chi Minh City, Vietnam
\\
{$^3$} Division of Nuclear Physics, Advanced Institute of Materials Science, Ton 
Duc Thang University, Ho Chi Minh City, Vietnam \\
{$^4$} Faculty of Applied Sciences, Ton Duc Thang University, Ho Chi Minh City, 
Vietnam}

\begin{abstract}
{
The study of CNO cycle involves the examination of the proton radiative capture, or the \pg reactions below 2 MeV. The astrophysical $\mathcal{S}$ factor characterizing the \pg reaction is usually reduced to the electric dipole transition $E1$ from the scattering state to the bound state. In this work, the partial scattering and the single-particle bound wave functions in the reduced matrix element of the transition are obtained from the single self-consistent mean-field potential deduced from the Skyrme Hartree-Fock calculation. The astrophysical $\mathcal{S}$ factors of the \pg reactions in the CNO cycle were successfully reproduced. The self-consistent Hartree-Fock calculation from the discrete to the continuum is a promising approach for the microscopic analysis of the nucleon-induced reactions in nuclear astrophysics.
}
\end{abstract}
\date{\today}
\maketitle

\section{Introduction}
The compilation of charged-particle-induced thermonuclear reaction rates--Nuclear Astrophysics Compilation of REactions (NACRE) \cite{ang99} was recently updated \cite{xu13}. Among these thermonuclear reactions, the proton radiative capture \pg is important to the understanding of the CNO cycle dominating in stars that are 1.3 times heavier than the Sun. From the theoretical point of view, various models have been adopted in the study of \pg reactions at the very low energies, such as the phenomenological $R$-matrix method, the ``microscopic cluster models'', and the potential model (see Ref.~\cite{des20} for a recent review). The so-called potential model is the simplest approach that concentrates on the calculation of the dipole electric transition $E1$ from the scattering state to the bound state. One can focus on calculating the overlap integral of the partial scattering wave function and the single-particle (s.p.) bound wave function. In many previous works, the phenomenological nuclear potentials such as the Gaussian, Woods-Saxon, and even the square-well potentials have been used to obtain the s.p. wave functions, for instance \cite{xu13,dub11,hua10,rol73,tia18}. The microscopic nuclear potential was also applied. The folding model with the density dependent effective nucleon-nucleon interaction that is usually used for the scattering problem at the energy above 20 MeV was recently presented in Ref.~\cite{anh2021}. The theoretical calculation for the scattering problem in the \pg reactions, however, have been usually approached in the direction from high positive energies to very low positive energies. In this work, the scattering problem was approached by using the opposite direction, from the negative energy region to the very low positive energy: the self-consistent Hartree-Fock (HF) approximation for the continuum.

The HF approximation gives good descriptions not only for the nuclear s.p. bound state, but also the scattering state at low energies \cite{dov71,dov72,ber79,ber80}. The HF s.p. potential in the continuum plays the role of the real part of the optical potential. The method was recently updated and applied for the study of nucleon-nucleus elastic scattering up to 40 MeV using the Skyrme interaction \cite{miz14,hao15} or the Gogny interaction \cite{bla15}. Therefore, the HF calculation is appropriate for the study of \pg reaction at the very low energies related to the nuclear astrophysics.

In the present work, the self-consistent mean-field potential obtained from the HF calculation with SLy4 interaction \cite{cha98} was used to determine simultaneously the scattering wave function and the s.p. bound wave functions. The strength of the real optical potential was fine-tuned to reproduce the low energy resonances. The nuclear data of the \pg reactions of the CNO cycle including $^{12,13}$C($p,\gamma$), $^{14}{\rm N}$($p,\gamma$), and $^{16}{\rm O}$($p,\gamma$) were successfully reproduced. The $E1$ transitions not only to the ground state (g.s.) but also to the excited states are both analyzed.

\section{Method of calculation}
\subsection{Potential model for \pg reactions}
In the study of the $A(p,\gamma)B$ reactions at the energies below the Coulomb barrier, it is customary to use the energy dependent astrophysical 
$\mathcal{S}(E)$ defined as
\begin{equation} \label{Sfactor}
\mathcal{S}(E) = E\exp(2\pi \eta) \sigma(E),
\end{equation}
where $\eta$ is the Sommerfeld parameter that depends on the charges and the relative velocity of the system. In our cases, within the potential model, the $E1$ partial cross-section $\sigma_{J'}(E)$ from initial (scattering) states $J^{\pi}$ to a given final (bound) state $J'^{\pi'}$ can be written as Ref.~\cite{hua10}
\begin{align} \label{part. c.s.}
\sigma_{J'} (E) = \dfrac{2}{9}\dfrac{(2\pi)^3}{k^2} 
\dfrac{2(J'+1)}{(2S_A+1)(2s_p + 1)}\dfrac{1}{4\pi}\left(Z_{A} \frac{m_p}{m_{B}} 
- Z_p \dfrac{m_{A}}{m_{B}} \right)^2 S_F k_{\gamma}^3  \nonumber \\
\times\sum\limits_{J j \ell} 3(2j'+1)(2J +1) \times \left\{ \begin{matrix}
j & J & S_A \\
J' & j' & 1
\end{matrix} \right\}^2 \times  \left( \begin{matrix}
j' & 1 & j \\
1/2 & 0 & -1/2
\end{matrix} \right)^2 \times \mathcal{I}^2(E).
\end{align}
$S_A$ and $s_p$ are the spin of the target and the incident proton. The quantity in the first parentheses is the effective charge number depending on the nucleus masses $m$ and charges $Z$. $\bm j = \bm \ell + \bm s_p$ and $\bm j' = \bm \ell' + \bm s_p$ are the total angular momenta of proton where $\ell$ and $\ell'$ are the relative orbital angular momenta of the entrance and exit channel, respectively. $k = E/{\hbar c}$ is the incident proton wave number at the bombarding energy $E$. The photon wave number is
\begin{equation}
 k_\gamma = \frac{E - (E_b + E_x)}{\hbar c},
\end{equation}
in which $E_x$ is the excitation energy of the daughter nucleus. $E_b$ is the proton separation energy of the proton-core ($A+1$) system. $S_F$ is the spectroscopic factor that is finally adjusted to reproduce the experimental data \cite{rol73, ili04}.

In this simple model, $\mathcal{I}(E)$ in Eq.~(\ref{part. c.s.}) is the radial overlap integral of the (initial) scattering wave function $\chi_{\ell}(E,r)$ and the (final) bound wave function $\phi_\alpha(r)$ ($\alpha$ stands for the set $n'\ell'j'$) expressed as 
\begin{equation} \label{Int}
\mathcal{I}(E) = \int \phi_\alpha(r) \chi_{\ell}(E,r) r dr.
\end{equation}
The self-consistent mean-field potential in the HF approximation can simultaneously generate the bound wave function $\phi_\alpha(r)$ and the scattering wave function $\chi_{\ell} (E,r)$ \cite{dov71, dov72}. The method can be applied to improve the so-called potential model, and hence is named the bound-to-continuum potential model. The model for the study of astrophysical \pg reaction is, therefore, based on a consistent and microscopic calculation. 

\subsection{The bound-to-continuum potential model}
For the bound s.p. wave function, the HF s.p. equation with the Skyrme interaction \cite{vau72,bei75} is given by
\begin{equation} \label{HFeq}
\dfrac{\hbar^2}{2m^*(r)}\left[- \phi_\alpha''(r) + \dfrac{\ell'(\ell' + 
1)}{r^2} \phi_\alpha(r) \right] + V_{\rm HF}(r)\phi_\alpha(r) - \left[\frac{\hbar^2}{2m^*(r)}\right]' \phi_\alpha'(r) = \epsilon_\alpha \phi_\alpha(r), 
\end{equation}
where $\epsilon_\alpha$ are s.p. energies of bound states and $m^*(r)$ is the nucleon effective mass. The HF potential $V_{\rm HF}(r)$ contains the central (c), spin-orbit (s.o.) and Coulomb (Coul.) potential
\begin{equation}
 V_{\rm HF}(r) = V_{\rm c}(r) + V_{\rm Coul.}(r) + V_{\rm s.o.}(r) \bm \ell \cdot \bm s.
\end{equation}

For the partial scattering wave function, the connection between the HF potential and the real part of the nucleon-nucleus optical potential were given in Refs.~\cite{dov71,dov72}. The partial scattering wave function $\chi_\ell(E,r)$ is the solution of the equation
\begin{equation} \label{Eq_scatt}
\dfrac{\hbar^2}{2m} \left[ - \chi_\ell''(E,r) + \dfrac{\ell(\ell + 1)}{r^2} \chi_\ell(E,r) \right] + \mathcal{V}(E, r)\chi_\ell(E,r) = E \chi_\ell(E,r),
\end{equation}
where $\mathcal{V}(E, r)$ in Eq.~(\ref{Eq_scatt}) is the real part of the optical potential
\begin{equation}\label{OMP}
 \mathcal{V}(E, r) = \lambda_c\mathcal{V}_{\rm c}(E, r) + \mathcal{V}_{\rm 
Coul.}(r) + \mathcal{V}_{\rm s.o.}(r) \bm l \cdot \bm s.
\end{equation} 
The optical potential $\mathcal{V}(E, r)$ is related to the HF potential $V_{\rm HF}(r)$ as \cite{dov72}
\begin{eqnarray} \label{scatt.pot}
\mathcal{V} (E, r) = \dfrac{m^*(r)}{m} \left[V_{\rm HF}(r) + 
\dfrac{1}{2} \left(\dfrac{\hbar^2}{2m^*(r)} \right)'' - \dfrac{m^*(r)}{2\hbar^2} 
\left[\left(\dfrac{\hbar^2}{2m^*(r)}\right)'\right]^2 \right] + \left[1 - \dfrac{m^*(r)}{m}\right] E.
\end{eqnarray}
As the energy range of interest is at the very low energies (below $2.0$ MeV), only the real part of the optical potential is considered. The imaginary part, and higher-order terms are expected to be negligible \cite{des20}. The result, $\mathcal{S}(E)$, is strongly sensitive to $\chi_\ell(E,r)$. Therefore, the adjustable $\lambda_c$ is multiplied to the strength of the central optical potential in Eq.~(\ref{OMP}) to obtain a better scattering wave function for the description of the resonance because it plays an important role in the study.

For the bound s.p. wave function, we follow previous works such as Ref.~\cite{hua10}. The wave function $\phi_\alpha(r)$ in Eq.~\eqref{Int} is the solution of the Schr\"{o}dinger equation for the bound state with the HF potential. The energy eigenvalue is chosen as $E_b = Q$ with $Q$ being the $Q$-value of the reaction. 

\begin{figure}[bht]
	\includegraphics[width=1.0\textwidth]{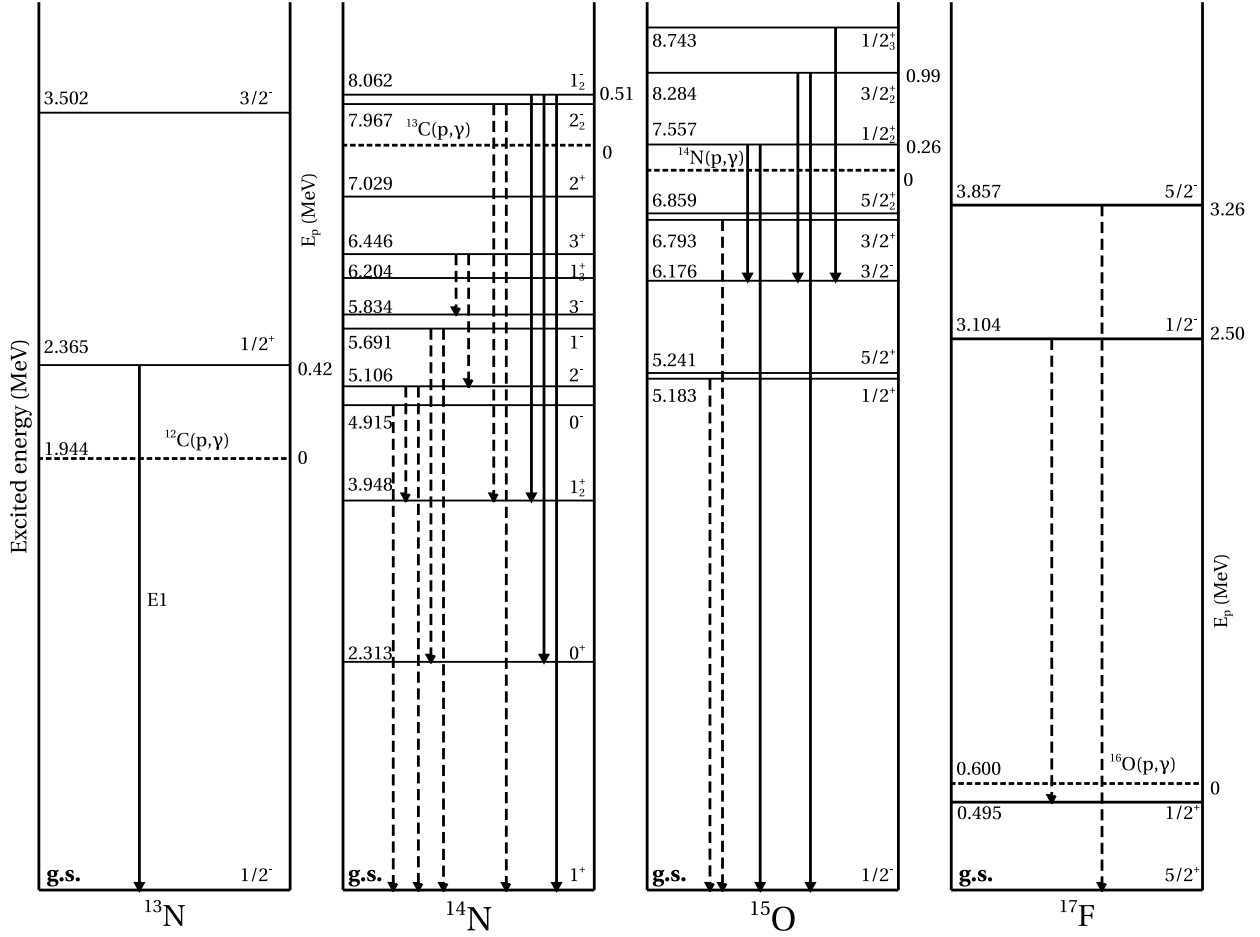}\vspace*{-0.0cm}
	\caption{Low excited states of $^{13}{\rm N}$, $^{14}{\rm N}$, $^{15}{\rm O}$, and $^{17}{\rm F}$ around the proton threshold (the dashed lines). The solid arrows show the transitions $E1$ to the bound states that are considered. The dashed arrows are the other possible $E1$ transitions. Irrelevant states are omitted for clarity. \label{lv}}
\end{figure}

Level schemes of $^{13}{\rm N}$, $^{14}{\rm N}$, $^{15}{\rm O}$ and $^{17}{\rm F}$ around the proton threshold (dash-line) are shown in Fig.~\ref{lv} using the data from \cite{sel91,til93}. The energy of the threshold is the $Q$-value at which the kinetic energy of the incident proton is zero. About 1 MeV above the threshold is the relevant energy region. All $E1$ transitions are indicated by the arrows in Fig.~\ref{lv}, but only the solid arrows are considered.

At low-energy scattering, the number of partial waves is limited. Furthermore, because of the choices for the bound states and the selection rule $\ell - \ell' = \pm 1$, only one or two partial scattering wave functions contribute to the calculation. However, the partial-wave analysis is complicated because of different $J^\pi$ of the entrance channel. Therefore, only dominant contributions are considered. Table~\ref{MainTable} shows the properties of scattering states and bound states of dominant contributions.

Consequently, the single HF potential $V_{\rm HF}(r)$ and $m^*(r)$ can determine simultaneously the bound and scattering wave functions. The HF potential $V_{\rm HF}(r)$ and the effective mass $m^*(r)$ are obtained from a Skyrme-HF program that is now widely used such as the program given in Ref.~\cite{col13}.

\section{Results and discussion}
\begin{table}
\centering
\caption{The main configurations that contribute in the calculation. 
Scaling factor $\lambda_c$ and spectroscopic factor $S_F$ are obtained with the SLy4 interaction. \label{MainTable}} 
\begin{tabular}{|c|l|crcc|cr|r|r|}
\hline \hline \toprule
No.	&	Reactions	&	$J^\pi$	&	$\ell$	&	$\lambda_c$	&	$J'^{\pi'}$ &	$Q$-value	&	$S_F$ &	$S_F$ [5]	& $S_F$ [2]	\\	\hline
1	&	$^{12}$C($p,\gamma$)$^{13}$N	&	$1/2^+$	&	$s$	&	1.15	&	$1/2^-$ &	1.944	&	0.36	&	0.35	&	0.33	\\	\hline
2	&	$^{13}$C($p,\gamma$)$^{14}$N	&	$1^-$	&	$s$	&	1.03	&	$1^+$	&	7.550	&	0.27	&	0.15	&	0.28	\\	
3	    &		    &	$1^-$	&	$d$	&	1.30	&	$1^+$	&	7.550	&	0.27	&	-	&	-	\\	\hline
4	&	$^{13}\mathrm{C}(p,\gamma)^{14}\mathrm{N}^*$ ($2.313$ MeV) 	&	$1^-$	&	$s$	&	1.03	&	$0^+$	&	7.550	&	0.08	&	-	&	0.027	\\	
5	&		&	$1^-$	&	$d$	&	1.30	&	$0^+$	&	7.550	&	$7 \times 10^{-3}$	&	-	&	-	\\	\hline
6	&	$^{13}\mathrm{C}(p,\gamma)^{14}\mathrm{N}^*$ ($3.948$ MeV) 	&	$1^-$	&	$s$	&	1.03	&	$1^+$	&	7.550	&	0.4	&	-	&	0.28	\\	
7	&		&	$1^-$	&	$d$	&	1.30	&	$1^+$	&	7.550	&	0.4	&	-	&	-	\\	\hline
8	&	$^{14}$N($p,\gamma$)$^{15}$O	&	$1/2^+$	&	$s$	&	1.08	&	$1/2^-$ &	7.297	&	$2 \times 10^{-3}$	&	-	&	$3.5 \times 10^{-5}$	\\	
9	&		&	$1/2^+$	&	$d$	&	1.32	&	$1/2^-$	&	7.297	&	0.02	&	-	&	-	\\	
10	&		&	$3/2^+$	&	$d$	&	1.28	&	$1/2^-$	&	7.297	&	$5 \times 10^{-3}$	&	-	&	$3.4 \times 10^{-5}$	\\	\hline
11	&	$^{14}\mathrm{N}(p,\gamma)^{15}\mathrm{O}^*$ ($6.176$ MeV)	&	$1/2^+$	&	$s$	&	1.08	&	$3/2^-$	&	7.297	&	0.13	&	-	&	$2.4 \times 10^{-5}$	\\	
12	&		&	$3/2^+$	&	$d$	&	1.28	&	$3/2^-$	 &	7.297	&	0.03	&	-	&	$1.7 \times 10^{-5}$	\\	
13	&		&	$1/2^+$	&	$d$	&	1.26	&	$3/2^-$	 &	7.297	&	0.46	&	-	&	$5.0 \times 10^{-3}$	\\	\hline
14	&	$^{16}\mathrm{O}(p,\gamma)^{17}\mathrm{F}$	&	$3/2^-$	&	$p$	&	1.00	&	$5/2^+$	&	0.600	&	1.00 &	0.90	&	-	\\	\hline
15	&	$^{16}\mathrm{O}(p,\gamma)^{17}\mathrm{F}^*$ ($0.495$ MeV)	&	$1/2^-$	&	$p$	&	1.00	&	$1/2^+$	&	0.600	&	1.00 &	1.00	&	-	\\	\hline \hline
\end{tabular} 
\end{table}

\subsection{$^{12}$C($p, \gamma$)$^{13}$N}
The starting point of the CNO cycle competing with the $pp$ chain in the hydrogen combustion phase is $^{12}$C($p,\gamma$)$^{13}$N reaction. In Fig.~\ref{lv}, there is the resonance at $E_p = 0.42$ MeV above the threshold corresponding to the first excited state of $^{13}$N at $E_x = 2.365$ MeV ($1/2^+$). As the g.s. of $^{13}$N has $J'^{\pi'} = 1/2^-$, the possible entrance channels are $J^\pi = 1/2^+, 3/2^+$. In our assumption, the incoming proton is captured into the s.p. state $1p_{1/2}$. In the partial wave analysis, the corresponding scattering partial wave of $J^\pi = 1/2^+$ is the $s$-wave that is the main contribution (as given in Table~\ref{MainTable}). In our calculation, the $p$-wave corresponding to $J^\pi = 3/2^+$ is negligible. 

The resonance plays the important role as the calibration for the calculation. It is emphasized that while the bound state $\phi_{\alpha}(r)$ is fixed at a given bound energy, $\chi_{\ell}(E,r)$ is energy dependent. Therefore, the astrophysical factor $\mathcal{S}(E)$ is more sensitive to the scattering wave function than to the bound wave function. The resonance is strongly sensitive to the scaling factor $\lambda_c$. Consequently, to generate the peak of the resonance at the given energy, $\lambda_c$ is fine-tuned to be $1.15$ in this case (Table~\ref{MainTable}). The value of $S_F$ is in agreement with the result in Ref.~\cite{hua10}.

\begin{figure}[bht]
	\includegraphics[width=0.6\textwidth]{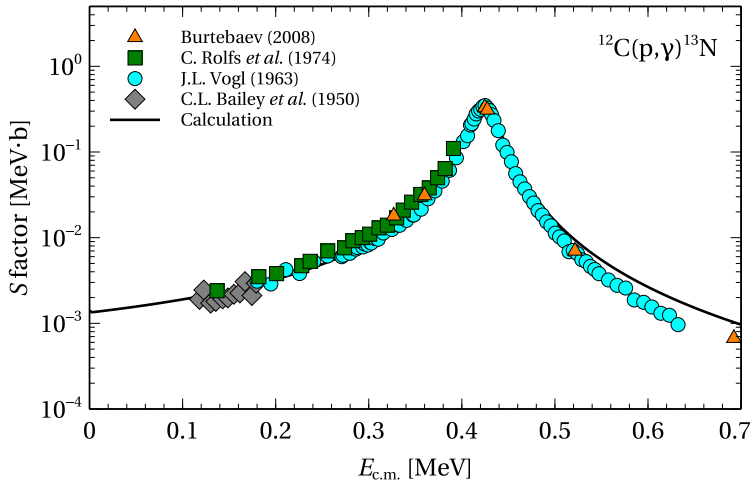}\vspace*{-0.5cm}
	\caption{The results for the reaction $^{12}$C($p,\gamma$)$^{13}$N. The experimental data is taken from Refs.~\cite{bai50,hal50,vog63,rol74}. \label{C12}}
\end{figure}

\subsection{$^{13}$C($p,\gamma$)$^{14}$N and $^{13}$C($p,\gamma$)$^{14}$N$^*$}
After the nucleus $^{13}$N produced from $^{12}$C\pg undergoes the beta plus decay, the next reaction of the CNO cycle is the $^{13}$C($p,\gamma$)$^{14}$N reaction. It plays a key role for nuclear energy production in massive stars and control the buildup of $^{14}$N. As a consequence, the $^{12}{\rm C}/^{13}{\rm C}$ ratio is reduced \cite{nes01}. This abundance ratio is one of important ratios for the measurement of stellar evolution and nucleosynthesis. In the same manner as for the reaction $^{12}$C\pg, there is the resonance at $0.51$ MeV corresponding to the $1^-$ excited state at $8.062$ MeV in the $^{14}$N level scheme (Fig.~\ref{lv}). 

\begin{figure}[bht]
\includegraphics[width=0.6\textwidth]{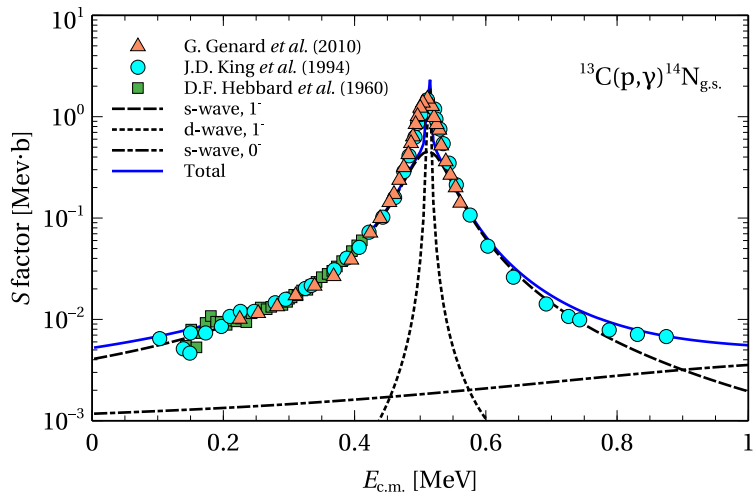}\vspace*{-0.5cm}
\caption{The $\mathcal{S}$ factor of $^{13}$C($p,\gamma$)$^{14}$N reaction. The dash line is the $s$-wave with $J^\pi = 1^-$ that is the main contribution. The dotted line is the $d$-wave with $J^\pi = 1^-$ that is the additional narrow peak. The dash-dotted line, $s$-wave with $J^\pi = 0^-$, is analyzed to improve the tail of the resonance. The solid line is the total calculation. The data are taken from Ref.~\cite{kin94}.} 
\label{C13GS}
\end{figure}

The $s$-wave is dominating in the partial wave analysis. The resonance is reproduced with the scaling factor $\lambda_c$ being $1.03$ and the spectroscopic factor $S_F$ being $0.27$. However, as same as the result of most of the previous works \cite{hua10}, the potential-model calculation can only produce the position of the peak at $0.51$ MeV (Fig.~\ref{C13GS}). The peak is lower than the value of the measurements of King \textit{et al.} \cite{kin94}. The possible calculation that can be done to improve the result is by taking into account the narrow resonance caused by the $d$-wave that usually was unnoticed in the previous works \cite{Kho20}. For the $d$-wave, a resonance appears with $\lambda_c = 1.30$ and $S_F = 0.27$. The difference between $S_F$ of $s$-wave and $d$-wave is not significant. However, their value of $\lambda_c$ are different. The explanation is that the $d$-wave is affected by the spin-orbit potential while $s$-wave is of course not. In our calculation, $\lambda_c$ is the scaling parameter only for the central optical potential while the spin-orbit potential is kept unchanged.

\begin{figure}[bht]
\includegraphics[width=0.9\textwidth]{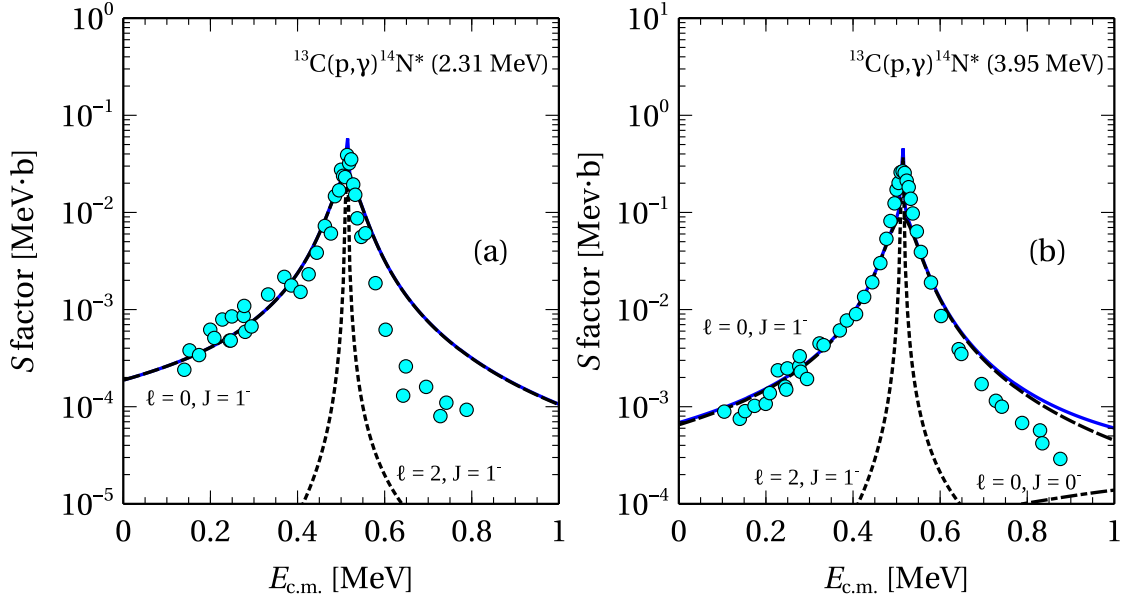}\vspace*{-0.5cm}
\caption{The same as Fig.~\ref{C13GS}, but for the transitions to the excited states, $^{13}$C($p,\gamma$)$^{14}$N$^*$, at 2.31 MeV (a) and 3.95 MeV (b). The data is taken from Ref.~\cite{kin94}.} 
\label{C13_Ex}
\end{figure}

Furthermore, a slightly better description for the tail of the resonance is reproduced by taken into account the contribution of the $s$-wave of $J^\pi = 0^-$ that is calibrated by the resonance at 8.776 MeV (the dash-dotted lines with $\ell =0, J^\pi = 0^-$ in Fig.~\ref{C13GS}). In the energy level scheme of $^{14}$N, there is a $2^+$ state at $7.967$ MeV. The data of the $E1$ transitions to $0^+$ (2.31 MeV) and $1^+$ (3.59 MeV) excited states are also reproduced in the same calculation for the transition to the g.s. (Fig.~\ref{C13_Ex}).

Note that for $^{13}$C case, our result is different from the value of Ref.~\cite{hua10}, but it is close to the value of $S_F$ in Ref.~\cite{anh2021}. The difference in the value of $S_F$ between Ref.~\cite{hua10} and Ref.~\cite{anh2021} was already discussed in Ref.~\cite{anh2021}.

\subsection{$^{14}\mathrm{N}(p,\gamma)^{15}\mathrm{O}$ and $^{14}\mathrm{N}(p,\gamma)^{15}\mathrm{O}^*$}
\begin{figure}[bht]
	\includegraphics[width=0.8\textwidth]{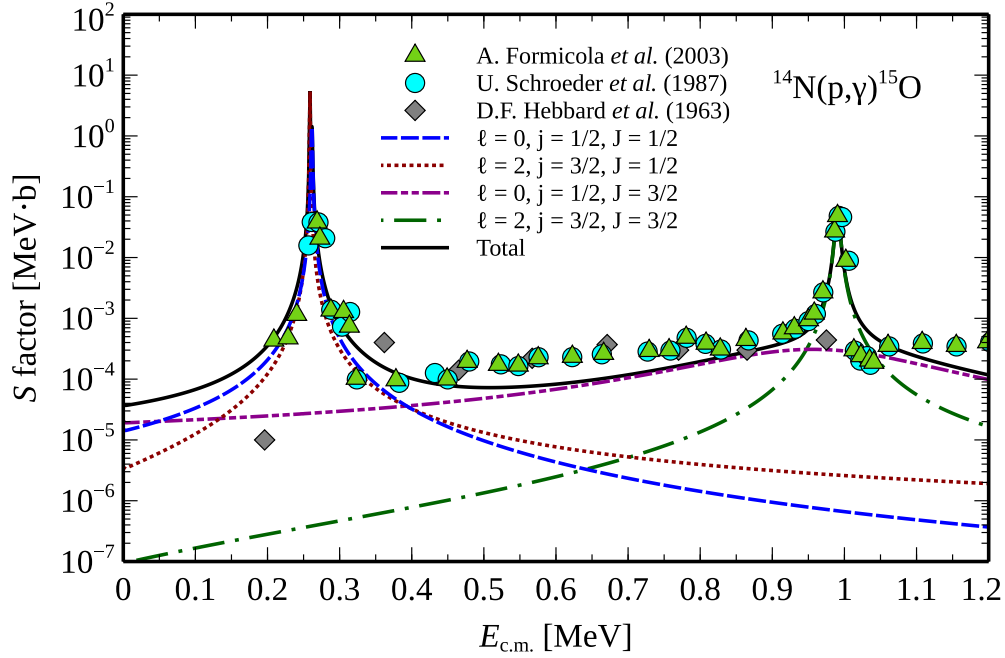}\vspace*{-0.5cm}
	\caption{The partial wave analysis of the reaction $^{14}$N($p,\gamma$)$^{15}$O. The experimental data are taken from Refs.~\cite{heb63,sch87,for03}.} \label{N14-full}
\end{figure}

The next reaction in the CNO cycle is the $^{14}$N($p,\gamma$)$^{15}$O that is the slowest reaction and thus controls the energy generation. The g.s. of $^{15}$O has $J'^{\pi'} = 1/2^-$. The calculation can be calibrated by two resonances at $0.26$ MeV and $0.99$ MeV corresponding to the excited states at $7.56$ MeV ($1/2^+_2$) and $8.28$ MeV ($3/2^+_2$), respectively. The partial wave analysis is shown in detail in Fig.~\ref{N14-full} for this reaction. At the first resonance, the scattering wave functions taken into account are $s$-wave and $d$-wave. The resonance is produced by the $s$-wave and $d$ wave with $\lambda_c = 1.08, S_F = 2 \times 10^{-3}$ and $\lambda_c = 1.32, S_F = 0.02$, respectively; using the SLy4 interaction. The second resonance is caused by the $d$-wave when the two parameters are $\lambda_c = 1.28$ and $S_F = 5 \times 10^{-3}$. The $s$-wave ($\ell = 0, j = 1/2, J = 3/2$) is also analyzed. It contributes to the background and slightly improves the result. Fig.~\ref{N14-full} shows that two resonances are well-reproduced in comparison with the experimental data \cite{heb63,sch87,for03}. The background lines (from 0.4 MeV to 0.9 MeV) are also well reproduced as the result from the sum of tails of resonances (Fig.~\ref{N14-full}).

For the $E1$ transition to the $3/2^-$ excited state of $^{15}$O at 6.176 MeV, there are three resonances below 1.5 MeV (Fig.~\ref{lv}) including two resonances in the previous case and one additional resonance $1/2^+_3$ at 8.74 MeV. The partial waves that are the main contributions to each resonance are given in Table~\ref{MainTable}. The same scaling factors $\lambda_c$ for the first two resonances are, of course, the same as that of the transition to the g.s.. For the third resonance, $\lambda_c$ is $1.26$, and $S_F$ is $0.46$.
\begin{figure}[bht]\vspace*{0cm}
	\includegraphics[width=0.8\textwidth]{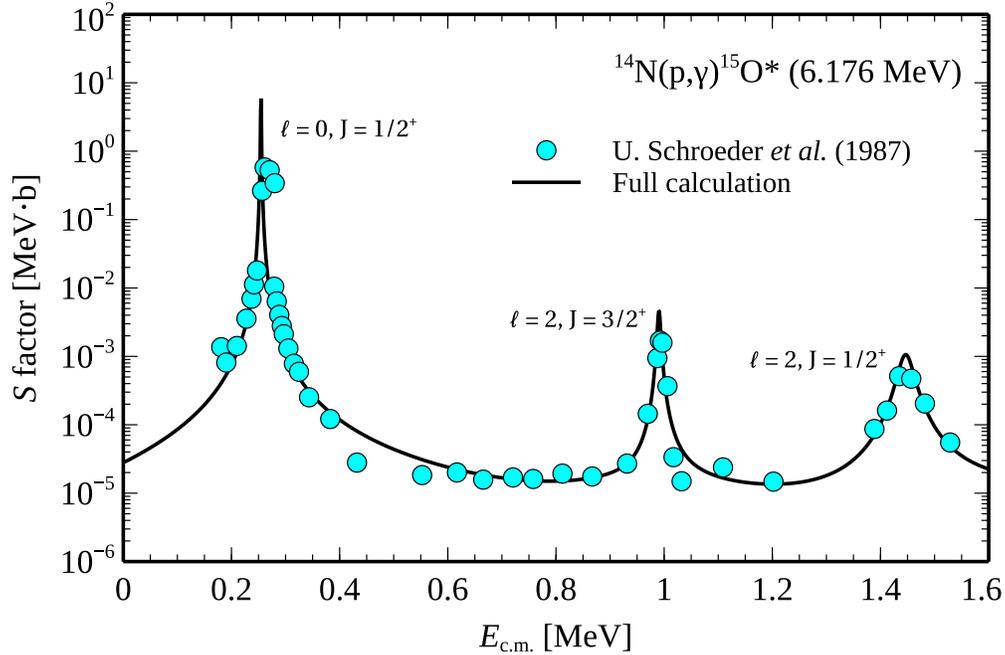}\vspace*{-0.5cm}
	\caption{The result for $^{14}\mathrm{N}(p,\gamma)^{15}\mathrm{O}^*$ (6.176 
MeV) compared to the experimental data from Ref.~\cite{sch87}.} 
\label{N14-ex}
\end{figure}

\subsection{$^{16}$O\pg$^{17}$F and $^{16}$O\pg$^{17}$F*}
Finally, consider the $^{16}$O\pg$^{17}$F reaction where the reaction rate sensitively affects the ratio $^{16}{\rm O}/^{17}{\rm O}$ predicted by models of massive stars \cite{ili08}. Noted that it has slowest reaction rate in the CNO cycle, as there is no resonance in the astrophysical energy (below 1 MeV) (see Fig.~\ref{lv}). The $E1$ transitions to the g.s. of $^{17}{\rm F}$ with $J'^{\pi'} = 5/2^+$ and to the excited state at 0.495 MeV $J'^{\pi'} = 1/2^+$ are considered. 

\begin{figure}[bht]
	\includegraphics[width=0.8\textwidth]{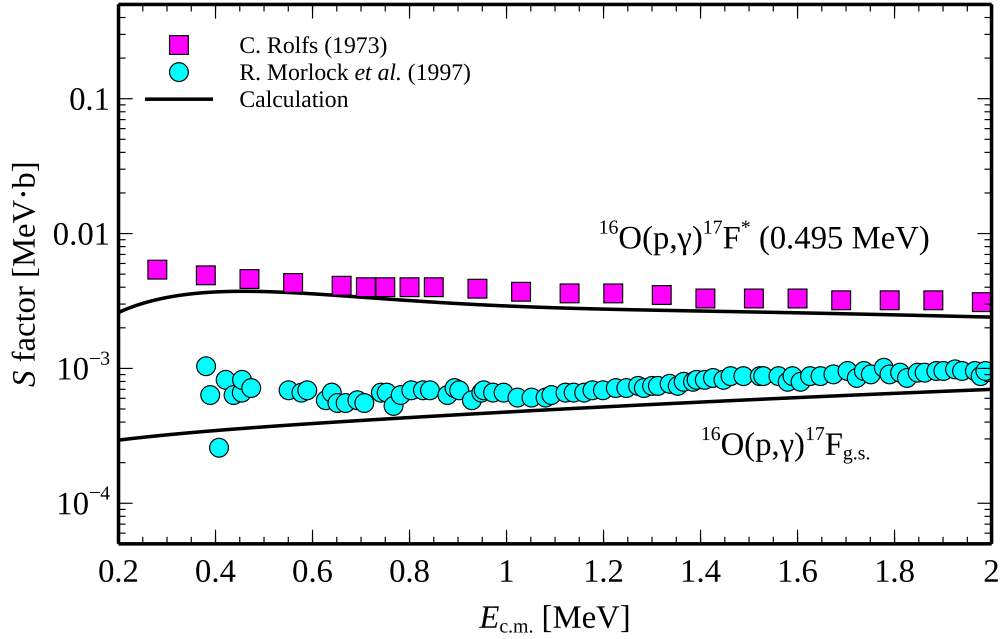}\vspace*{-0.5cm}
	\caption{The calculations for the reaction $^{16}$O($p,\gamma$)$^{17}$F. The experimental data are taken from Refs.~\cite{mor97,rol73}.} 
\label{O16}
\end{figure}

For the transition to the g.s., as given in Table~\ref{MainTable}, the proton is captured into the $1d_{5/2}$ s.p. state, therefore, in the partial wave analysis, the $p$-wave is the main contribution (see Table~\ref{MainTable}). The other possible contributions, for example, $f$-wave with $J^\pi = 5/2^-, 7/2^-$ are just less than $0.1\%$ in comparison with the main contribution. The result in Fig.~\ref{O16} shows that the calculation reproduces the experimental data \cite{mor97}. It is emphasized that $\lambda_c$ is unity. The scaling factor $\lambda_c$ is unity because the HF to continuum gives a good result for the proton elastic scattering from $^{16}$O at low energy \cite{dov71}.

For the transition to the $1/2^+$ excited state of $^{17}$F at 0.495 MeV, the incident proton is added into the $2s_{1/2}$ s.p. state. It means that the excited state of $^{17}$F$^*$ is simply assumed to be built from one proton excited to $2s_{1/2}$ from $1d_{5/2}$, and the difference in the energy of the two states is 0.495 MeV corresponding to the excited energy of $^{17}$F$^*$, $1/2^+$. The contributions come from the $p$-wave with $J^\pi = 1/2^-, 3/2^-$. The $J^\pi = 1/2^-$ is dominating, because it is supposed to be the tail of the narrow resonance corresponding to the $1/2^-$ state at 3.104 MeV close to the energy region of interest (Fig.~\ref{lv}). The scaling factor $\lambda_c$ is, of course, the same as the case of the transition to the g.s.. According to the assumption for the excited state of $^{17}$F, with $S_F$ being $1.0$, the data \cite{rol73} is successfully reproduced as shown in Fig.~\ref{O16}.

\section{Conclusion}
The study of \pg reaction in the CNO cycle is approached microscopically and consistently by using the self-consistent mean-field method. Within the potential model, the bound and scattering wave functions can be obtained simultaneously from the single self-consistent mean-field potential. The approach can be applied to the transition not only to the g.s. but also to the excited states, and therefore can reproduce most of available experimental data for the astrophysical \pg reactions in the NACRE database. Strictly speaking, the pairing correction and deformation should be taken into account in the calculation, except for the case of $^{16}$O. Our results show that the HF calculation is a reasonable approach for the \pg reactions in the study.

\section*{Acknowledgements}
We would like to thank Prof. Dao T. Khoa, Prof. N. Auerbach, and Dr. A. Idini for discussions. The work is supported by Vietnam National Foundation for Science and Technology Development (NAFOSTED).

\bibliography{pgamma_revised}
\end{document}